\documentclass[dvips]{acta}
\usepackage{supertabular,lscape,epsfig}
\usepackage{amssymb}
\usepackage{amsmath}

\usepackage{color}
\definecolor{mmr}{rgb}{0.7,0.0,0.2}

\SetPages{0}{0}

\SetVol{69}{2019}

\begin{document}

\begin{Titlepage}
\Title{Hen\,3-160 -- the first symbiotic binary with Mira variable S star}
\Author{C. Ga{\l}an$^1$, J. Miko{\l}ajewska$^1$, B. Monard$^2$, K. I{\l}kiewicz$^1$, D. Pie\'nkowski$^1$, and M. Gromadzki$^3$}
{$^1$Nicolaus Copernicus Astronomical Center, Polish Academy of Sciences, Bartycka 18, PL-00-716 Warsaw, Poland\\
e-mail: cgalan@camk.edu.pl\\
$^2$Kleinkaroo Observatory, Calitzdorp, Western Cape, South Africa\\
$^3$Warsaw University Astronomical Observatory, Al. Ujazdowskie 4, PL-00-478, Warsaw, Poland
}

\Received{October 9, 2018}
\end{Titlepage}

\Abstract{Hen\,3-160 is reported in Belczy\'nski et al.'s  (2000) catalog as
a symbiotic binary system with M7 giant donor.  Using $V$- and $I$-band
photometry collected over 20 years we have found that the giant is a Mira
variable pulsating with 242.5-day period.  The period-luminosity relation
locates Hen\,3-160 at the distance of about 9.4\,kpc, and its Galactic
coordinates ($l=267.7^{\circ}$, $b=-7.9^{\circ}$) place it $\sim$1.3\,kpc
above the disc.  This position combined with relatively high proper motions
(pm$_{\rm{RA}}=-1.5$\,mas\,yr$^{-1}$, pm$_{\rm{DEC}}=+2.9$\,mas\,yr$^{-1}$;
Gaia\,DR2) indicates that Hen\,3-160 has to be a Galactic extended
thick-disc object.  Our red optical and infrared spectra show the presence
of ZrO and YO molecular bands that appear relatively strong compared to the
TiO bands.  Here we propose that the giant in this system is intrinsic S
star, enriched in products of slow neutron capture processes occurring in
its interior during an AGB phase which would make Hen\,3-160 the first
symbiotic system with Mira variable S star.}
{Stars: abundances - Stars: binaries: symbiotic - Stars: evolution - Stars:
late-type - Stars: individual: Hen\,3-160}

\section{INTRODUCTION}

The star Hen\,3-160 (other designations: SS73\,9, WRAY\,15-208,
Schwartz\,1; 2MASS\,08245314-5128329) is a
symbiotic binary included in the Allen (1984), Kenyon (1986), and Belczy\'nski et al. 
(2000) catalogs of symbiotic stars (SySt).

According to the Astrophysics Data System (ADS) it was first included in the
list of H$\alpha$ emission objects in the southern Milky Way by Wray (1966,
table 15), however, it was not bracketed together with zirconium 'S' stars. 
Sanduleak \& Stephenson (1973) included the object in the list of stars in
the Southern Milky Way with strong emission lines.  They noted
Z-And-like emission line spectrum with sharp He\,II\,(4686\,\AA) and
strong hydrogen emissions with an addition of weak He\,I and forbidden
nebular lines and proposed it to be a candidate for SySt.  Its SySt nature
was soon confirmed by Allen (1978) who described the system as very
high-excitation SySt with very strong He\,II (4686\,\AA), H$\beta$, [Fe
VII], [Ca VII], and [Ar X] emission lines.  Henize (1976) included this
object in his catalog of emission-line stars based on its H$\alpha$
emission.  Schwartz (1977) also noted the presence of numerous Balmer
emission lines in Hen\,3-160 spectrum.

There were attempts to detect the object in X-ray or radio waves, but no
positive detection was obtained in either case, not in X-rays with ROSAT
(Bickert et al.  1996), nor in radio domains (Wright \& Allen 1978; Wendker
1995).  

Based on analysis of 2MASS photometry by means of
($J-H$)\,--\,($H-K_{\rm{S}}$) diagram, Phillips (2007) confirmed the
classification of this star to the S-type{\footnote{Historical precedent has
resulted in two uses of S for late-type stars.  S stars have an
overabundance of s-process elements and typically C$/$O $\sim$ 1.  S-type
('stellar') symbiotic binaries are systems that lack dust.}} SySt.  Based on
near-infrared colors, Allen (1982) divided all SySt into two main classes:\\
({\sl{i}}) most ($\sim 80\%$) belong to a group which near-infrared spectra
are generally dominated by the cool star's photosphere, and are
indistinguishable from ordinary late-type giants (designated 'S' -- for
stellar);\\
({\sl{ii}}) the remaining ($\sim 20\%$) SySt exhibit the presence of additional
emission due to circumstellar, thick dust shells (D-type).  From near-$IR$
photometric monitoring, it is known that D-type SySt show large amplitude
variations that result from the presence of Mira variables -- since they
must accommodate the Mira with its dust shell, the orbital periods in these
cases should be as long as decades.

Although Hen\,3-160 was included in a number of works on large samples of
stars characterized with emission features, it has been never subjected so
far to any detailed studies and a little is known about the parameters of
its components.  In this work, we present new observations collected over
two decades which enabled us to reveal its very interesting nature.  In
particular, using our long-term $V$- and $I_{\rm C}$-band photometry and
optical spectra together with all available near-$IR$ measurements we show
here that the giant in this system is intrinsic S star, enriched in products
of slow neutron capture processes (s-process) occurring in its interior
during an AGB phase.  This makes Hen\,3-160 probably the first known SySt
with Mira variable S star.

\section{Observations and Reductions}

Optical spectra were obtained with SpUpNIC spectrograph (Crause et al. 
2016; Crause et al.  2018 -- in preparation) operated on 1.9\,m 'Radcliffe'
telescope in Sutherland (South African Astronomical Observatory) during two
observing seasons in March 2016 and October 2017.  Two gratings (5 and 11)
were used to obtain spectra in regions around the H$\alpha$ line ($\lambda$
$\sim$ 5950 -- 7150\,\AA, R $\sim$ 3900), and around the Ca\,II triplet
($\lambda$ $\sim$ 7100 -- 9700\,\AA, R $\sim$ 2200), respectively. 
Additionally, we have an older spectrum obtained in November 2005 with the
same telescope but equipped with previously mounted Cassegrain spectrograph
(SpCCD).  This spectrum covers whole optical region ($\sim$ 3800 --
7200\,\AA) in significantly lower resolution (R $\sim$ 1000).

\MakeTable{@{}c@{\hskip 2mm}c@{\hskip 2mm}c@{\hskip 2mm}c@{\hskip 2mm}c@{\hskip 2mm}c@{\hskip 2mm}c@{\hskip 2mm}c@{}}{12.0cm}{
Journal of spectroscopic observations obtained with 1.9\,m telescope at SAAO with information on the acquisition and exposure 
times, spectral ranges, resolution, spectrograph with which they were acquired, and the measured radial velocities.
}
{\hline
Date           & HJD        & Phase$^a$ & Spectral Reg. & Spectrograph & Exp. time  & Res.\,pow.                 & RV         \\
yyyy\,mm\,dd   & $-2450000$ &           &  $[$\AA$]$    &              & $[$sec.$]$ & $\lambda$/$\Delta \lambda$ & $[$km/s$]$ \\
\hline
  2005\,11\,20 & 3694.573 & 0.031 & 3810--7220 & SpCCD   & 3 $\times$ 300 & $\sim$ 1000 & $+109$ \\
  2016\,03\,16 & 7464.411 & 0.575 & 5950--7050 & SpUpNIC & 5 $\times$ 300 & $\sim$ 3900 & --     \\
  2016\,03\,20 & 7468.460 & 0.592 & 7150--9700 & SpUpNIC & 5 $\times$ 400 & $\sim$ 2200 & $+114$ \\
  2017\,10\,22 & 8048.614 & 0.984 & 6050--7150 & SpUpNIC & 12 $\times$ 60 & $\sim$ 3900 & $+116$ \\
  2017\,10\,30 & 8056.612 & 0.017 & 7100--9700 & SpUpNIC & 5 $\times$ 200 & $\sim$ 2200 & $+109$ \\
\hline
\multicolumn{8}{p{12.0cm}}{$^a$ Pulsation phase according to ephemeris: JD$_{\rm{max}}$= 2457810.0 + 242.53 $\times$ E
(see: Section\,3)}
}

\noindent All the spectra were reduced, extracted, wavelength calibrated,
heliocentric corrected, and calibrated into relative fluxes by standard
procedures using {\sl{IRAF}}\,\footnote{IRAF is distributed by the National
Optical Astronomy Observatories, which are operated by the Association of
Universities for Research in Astronomy, Inc., under a cooperative agreement
with the National Science Foundation.} packages.  The Journal of all
spectroscopic observations is shown in Table\,1.

\begin{figure}[htb]
\includegraphics{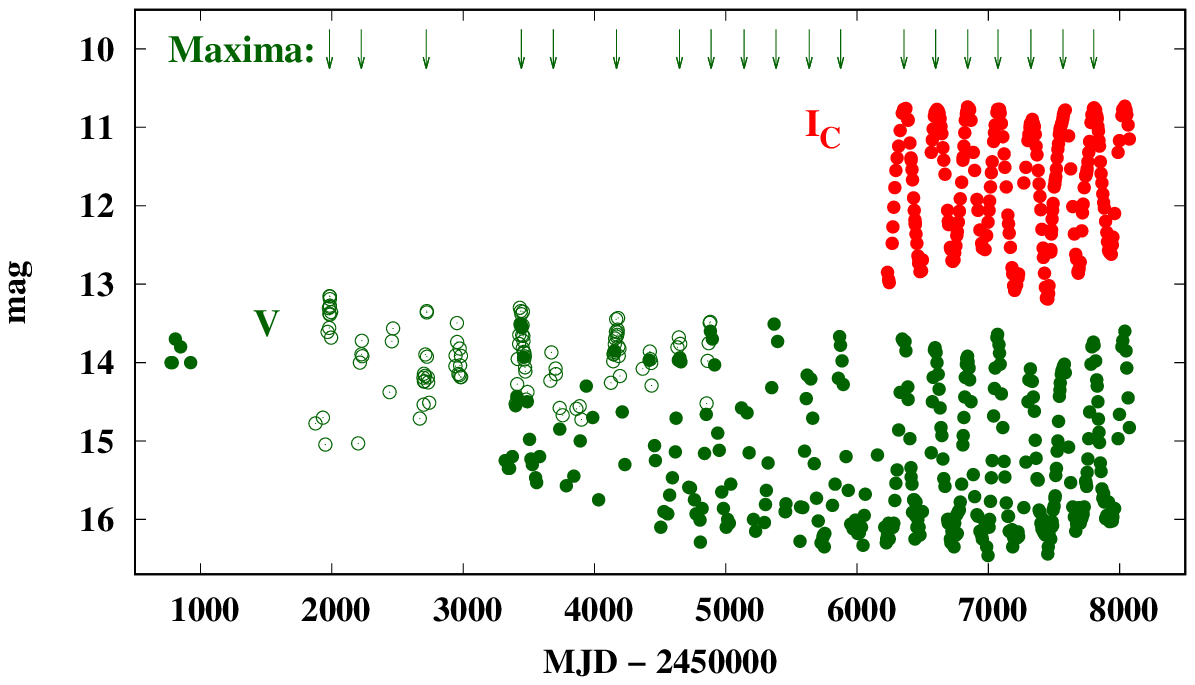}
\FigCap{$V$ and $I_{\rm{C}}$ light curves of Hen\,3-160.  Observations
obtained at the Kleinkaroo Observatory (solid points) are complemented with
the ASAS data (circles).  With arrows are shown the moments of pulsation
maxima identified and measured in the $V$ light curve (Section\,3).}
\end{figure}

%PHOTOMETRIC
Photometric, optical observations of Hen\,3-160 in $V$ filter were collected
during a 20-year period -- since November 22, 1997 (JD\,2450775.4) till
November 17, 2017 (JD\,2458074.57) -- with a 35\,cm Meade RCX400 telescope
in Kleinkaroo Observatory equipped with an SBIG ST8-XME CCD camera.  Since
November 1, 2012 (JD\,2456232.60) $I_{\rm{C}}$ filter was also included for
the monitoring.  Each single data point is the result of several individual
exposures, that were calibrated (dark-subtraction and flat-fielding) and
stacked selectively.  Magnitudes were derived from differential photometry
to nearby reference stars using the single image mode of AIP4 image
processing software.  Hen\,3-160 was also monitored with the All Sky
Automated Survey (ASAS, Pojma\'nski 1997) in $V$ photometric band since
November 27, 2000 (JD\,2451875.74) up to February 20, 2009 (JD\,2454882.77). 
$V$ and $I_{\rm{C}}$ light curves are shown in Figure\,1.

We also used a number of brightness measurements from the literature and
published catalogs (Appendix: Table\,7).  2MASS All-Sky Catalog of Point
Sources (Cutri et al.  2003) gives $JHK$ magnitudes obtained on
JD\,2451140.8321.  Kenyon et al.  (1988) collated IRAS photometry of SySt
providing values of fluxes (in Jy) for Hen\,3-160 at 12\,$\mu$m and
upper estimates for IRAS bands at a longer wavelength.  Using NASA/ IPAC
Infrared Science Archive\footnote{IRSA (NASA/IPAC Infrared Science Archive),
http://irsa.ipac.caltech.edu/frontpage/} we extracted infrared photometry
from new space infrared missions.  AKARI/IRC Point Source Catalogue contain
one measurement at 9\,$\mu$m.  From AllWISE Source Catalog we extracted
magnitudes in $W1$--$W4$ photometric bands.  Additionally, we extracted 55
photometric measurements in total from AllWISE Multiepoch Photometry Table
that were made around three dates $\sim$JD\,2455341, $\sim$JD\,2455528 and
$\sim$JD\,2455531 -- these are collected in Appendix: Table\,8.

\section{Pulsational variations of Mira star}

$V$-band photometric data set (complemented with ASAS data) was used for
Fourier's analysis which was performed using the discrete Fourier transform
method in the Period 04 program (Lenz \& Berger 2005).  The resulting power
spectrum is presented in Figure\,2.  We derived the value of pulsation
period (P$_{\rm{pul}}$) and time T$_0$ corresponding to pulsation maximum at
the zero epoch (E=0) with which the ephemeris for the maxima of stellar
pulsation can be written as follows: JD$_{\rm{max}}$= 2457813.8 + 242.53
($\pm$0.14) $\times$ E.  The resulting T$_0$ value
is artificially shifted -- it clearly does not fall on the pulsation
maximum -- presumably due to asymmetrical (sawtooth-like) shape
of a pulsational light curve.  Moreover, the Monte Carlo method implemented
in Period\,04 program to calculate errors results in severely underestimated 
error of T$_0$ -- the obtained value of $\pm0.03$ is
unrealistically small.\\
Phase dispersion minimalization (PDM) analysis
(Fig.\,2 -- bottom) gives period 244\,days.  In both cases of the power
spectrum and PDM similar period being a multiple of pulsation period is
obtained and some aliases are present as well as strong noise at low
frequencies.

\begin{figure}[htb]
\includegraphics{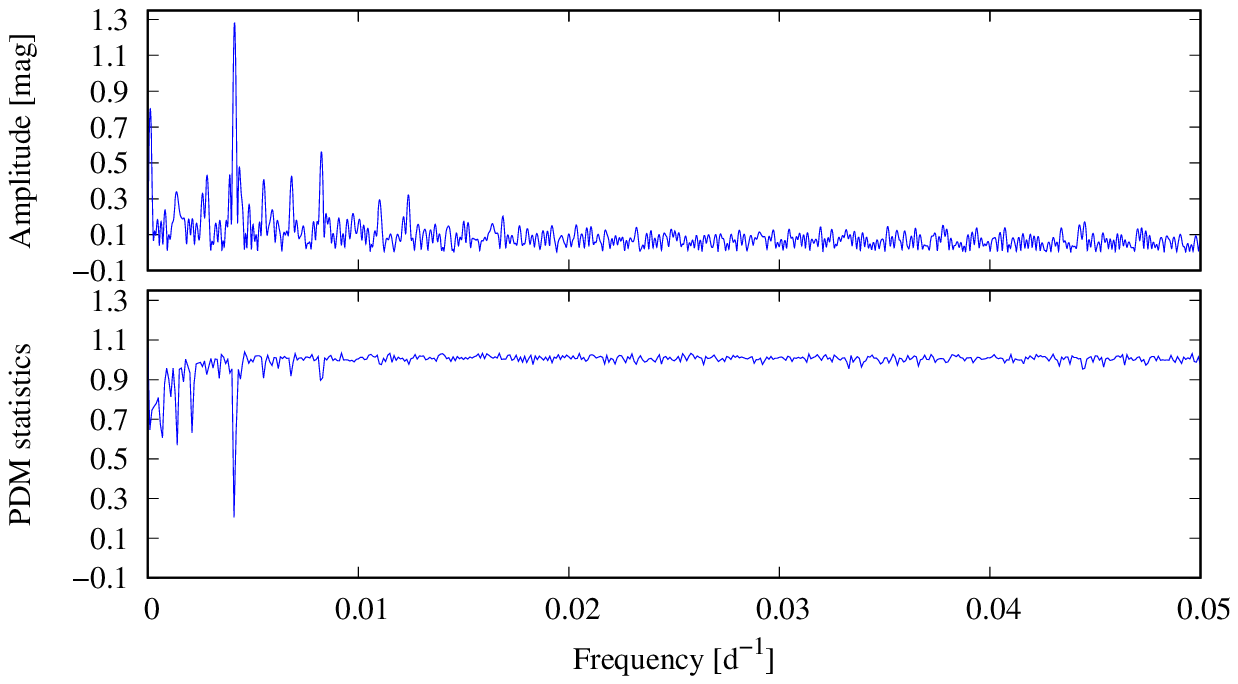}
\FigCap{Power spectrum ({\sl{top}}) and PDM window ({\sl{bottom}})
obtained using Fourier's analysis.}
\end{figure}

\MakeTable{cccc}{12.5cm}{Pulsation maxima and the $O-C$ residuals
calculated according to the initial ephemeris obtained with Fourier's method
(column 3) and the new ephemeris resulted from $O - C$ analysis (column 4).}
{\hline
Epoch no. &JD$-2450000$& $O-C$\,$[$d$]$ & $O-C$\,$[$d$]$ \\
\hline
$-24$     &  1982.57   &  $-10.47$    &  $-2.71$     \\
$-23$     &  2224.63   &  $-10.94$    &  $-3.35$     \\
$-21$     &  2718.95   &  $ -1.68$    &  $ 5.58$     \\
$-18$     &  3443.50   &  $ -4.72$    &  $ 2.04$     \\
$-17$     &  3686.23   &  $ -4.52$    &  $ 2.07$     \\
$-15$     &  4168.20   &  $ -7.61$    &  $-1.35$     \\
$-13$     &  4648.36   &  $-12.51$    &  $-6.59$     \\
$-12$     &  4888.43   &  $-14.97$    &  $-9.21$     \\
$-11$     &  5139.76   &  $ -6.17$    &  $-0.58$     \\
$-10$     &  5383.68   &  $ -4.78$    &  $ 0.64$     \\
$-9 $     &  5635.43   &  $  4.44$    &  $ 9.70$     \\
$-8 $     &  5874.88   &  $  1.36$    &  $ 6.45$     \\
$-6 $     &  6358.50   &  $ -0.08$    &  $ 4.68$     \\
$-5 $     &  6599.16   &  $ -1.95$    &  $ 2.64$     \\
$-4 $     &  6842.55   &  $ -1.09$    &  $ 3.33$     \\
$-3 $     &  7073.14   &  $-13.03$    &  $-8.78$     \\
$-2 $     &  7324.74   &  $ -3.96$    &  $ 0.13$     \\
$-1 $     &  7569.50   &  $ -1.73$    &  $ 2.19$     \\
$0  $     &  7803.14   &  $-10.62$    &  $-6.87$     \\
}
%-16      &  3947.81   &   14.53      &              \\	%Not used for timing analysis because of high (extreme) $O-C$ values
%1        &  8033.96   &  -22.33      &              \\	%Not used for timing analysis because of high (extreme) $O-C$ values

Since the Fourier's method was insufficient to obtain a realistic value for
time T$_0$, we carried out additional analysis of $O - C$ (observations minus
calculations) diagram performed on 19 moments of maxima identified in the
$V$ light curve (Figure\,1) which enable to get credible zero point of
the ephemeris.  The times of maxima were obtained by fit of the second
order polynomial (parabola) to the data and are listed in Table\,2 together
with $O - C$ residuals in respect to ephemeris obtained with use of
Fourier's method and in respect to the new ephemeris with improved T$_0$
point: JD$_{\rm{max}}$= 2457810.0 ($\pm$2.2 ) + 242.70 ($\pm$0.17) $\times$
E.  For the final ephemeris we adopt the zero point (T$_0$) resulting from
the $O - C$ analysis and the period resulting from Fourier's method which
uses all points from the light curve and thus should result in more accurate
value of the period.\\

JD$_{\rm{max}}$= 2457810.0 ($\pm$2.2 ) + 242.53 ($\pm$0.14) $\times$ E.\\

\noindent Both, $V$ and $I_{\rm{C}}$ light curves folded with pulsation
period according to the above ephemeris are shown in Figure\,3.

\noindent It is worth noting that all other observations -- both photometric
and spectroscopic -- in cases when a time of observations is known, show
changes correlated with the pulsation phase, eg.  WISE photometry in $W1$-
and $W2$-bands shows small amplitude variations in line with the trend of
brightness changes in the $V$ and $I$ light curves (Appendix:\,Table\,8).

\begin{figure}[htb]
\includegraphics{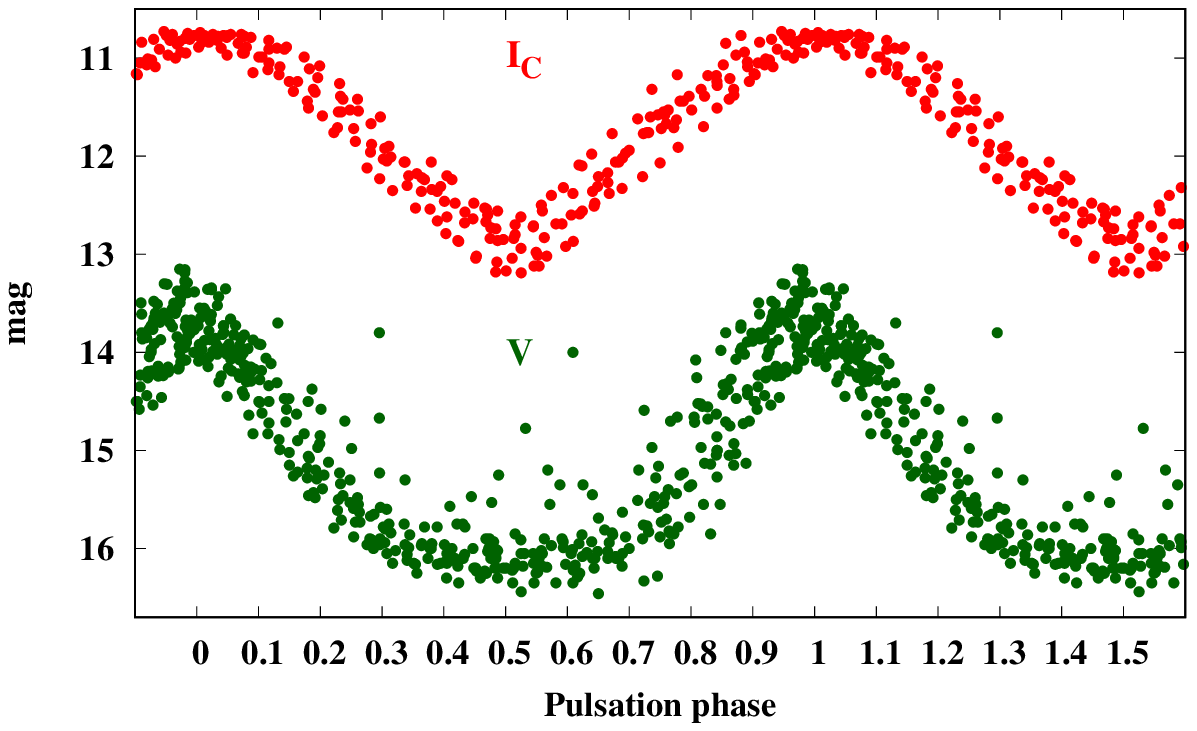}
\FigCap{$V$ and $I_{\rm{C}}$ light curves folded with pulsation period.}
\end{figure}

\section{Distance and position in the Galaxy}

Using known infrared magnitudes and relations binding them with parameters
of Mira stars, we can estimate the distance to Hen 3-160 and hence its
position in the Galaxy.  The period-luminosity relation for O-rich Galactic
Miras\\

M$_{\rm{K}} = -3.51 (\pm0.20) \times (\log{P} - 2.38) - 7.25 (\pm0.07)$ (Whitelock et al.  2008)\\

\noindent gives the absolute $K$ magnitude of the Mira in Hen\,3-160, M$_{\rm{K}}=-7.27\pm 0.07$ for
P$=242.5$\,d, with the error inferred from the Whitelock et al relation (note that the error in the period is very small error of order half of permille).  
Using near-$IR$ photometry from
2MASS and DENIS in $J$ and $K$ bands (Appendix: Table\,7) we obtain the
average $\overline{K}= 7.81 \pm 0.03$ and $(J-K) \sim 1.47$.  The intrinsic
period-color relation for Oxygen Miras\\

$(J-K)_0 = 0.71 (\pm0.06) \times \log{P} - 0.39 (\pm0.15)$ (Whitelock et al.  2000)\\

\noindent gives $(J-K)_0=1.30$ and hence E$_{J-K}\sim0.17$ what is more or
less consistent with the total Galactic E$_{J-K} = 0.26 \pm 0.02$ derived
with using the maps of the Galactic extinction by Schlafly \& Finkbeiner
(2011).  The reddening corrected $K$ magnitude is then $K_0=7.60\pm0.11$,
and the \textbf{distance to Hen\,3-160 is d$=9.4\pm1.4$\,kpc} what remains
in perfect agreement with the value of 9.5\,kpc estimated by Harries \&
Howarth (1996).  The Galactic coordinates ($b=-7.87^{\circ}$,
$l=267.6767^{\circ}$) would then result in $z=1.3\pm0.2$\,kpc which combined
with high proper motions ($\mu_{\alpha} cos{\delta} =
-1.504\pm0.079$\,mas\,yr$^{-1}$, $\mu_{\delta} =
+2.940\pm0.082$\,mas\,yr$^{-1}$) from Gaia DR2 (Gaia Collaboration 2016,
2018) suggests that Hen\,3-160 could be a Galactic halo object. 
Using M$_K=-7.27\pm 0.07$ and BC$_{\rm{K}} = 3.03 \pm 0.28$ we estimate
M$_{\rm{bol}} = -4.24 \pm 0.35$ and thus luminosity L $= 3960 (+1510/-1090)
L\odot$.

Gaia DR2 gives for Hen\,3-160 a negative parallax ($-0.043 \pm
0.040$\,mas\,yr$^{-1}$) with goodness-of-fit statistic parameter
{\sl{gofAL}} $\sim 23$, that indicates a very poor fit to the data. 
Bailer-Jones et al.  (2018) adopted a special method for estimating
distances from Gaia DR2 data, which allows obtaining this information even
with the use of negative values of parallaxes.  They obtained for Hen\,3-160
the distance 12.5\,kpc placed in the asymmetric confidence interval from 9.7
to 16.3\,kpc which overlaps with the distance derived here using the
M$_{\rm{K}}$ magnitude.  Our distance would correspond to the parallax
$0.106^{+0.019}_{-0.014}$\,mas\,yr$^{-1}$.  This value is by a factor of
$\sim30$ smaller than the proper motion from Gaia DR2 and is of the order of
its error.  Consequently, uncertainty in the parallax cannot bring more than
$\sim3-4\%$ of the proper motion value to the error of this parameter and is
irrelevant from the point of view of current considerations.

The Galactic coordinates, distance, and proper motions together with
systemic velocity ($\gamma = 112 \pm 2$\,km/s) estimated from measured
radial velocities (Table\,1) enabled us to estimate Galatic velocities
(U=$-150.3$; V=$-107.5$; W=$+3.3$\,km/s).  {\bf{The obtained values place the
system in the extended thick-disc}} (see eg.  Feltzing et al.  2003 --
figure\,1).

\section{Changes in spectra and temperature}

Based on TiO-bands strength in the near infrared region M{\"u}rset \& Schmid
(1999) estimated the spectral type M7 for the cool component.  An infrared
domain is particularly useful for studying properties of cool components in
SySt as they dominate this spectral range and contribution from the hot
component is usually negligible.  M{\"u}rset \& Schmid used spectra obtained
at JD\,2448696 and JD\,2449473, which correspond to pulsation phases $0.42
\pm 0.04$ and $0.62 \pm 0.04$ (according to our ephemeris), respectively. 
Our spectra covering this region (around Ca\,II triplet) are obtained at $\phi
= 0.017 \pm 0.010$ (October\,30,\,2017) and $\phi = 0.592 \pm 0.011$
(March\,20, 2016).  The first one is placed almost exactly at the pulsation
maximum, the second very close to pulsation minimum.  Therefore, it is
possible to evaluate the range of the Mira spectral type changes due to
pulsations by comparing these spectra with those of M-type giant standards. 
We have the spectra for above a dozen of such M giant spectroscopic
standards (Feast et al.  1990; Schulte-Ladbeck 1988) that were acquired
during the same seasons as those for Hen-3\,160 with the same instrument in
identical configuration and thus in exactly the same resolution.\\

\MakeTable{cccccccc}{12.5cm}{Spectral classification of Hen\,3-160 based on
the strength of TiO band heads in the spectra obtained with SpUpNIC
spectrograph (Grating no.  11).}
{\hline
Date   & Phase$^a$ & \multicolumn5c{TiO band head ($\lambda$ [\AA]).} &         \\
yyyy\,mm\,dd &           & 7589 & 8194 & 8432 & 8859 & 9209                 & Adopted \\
\hline
 2016\,03\,20   & $0.592 \pm 0.011$ & $>$M4.5 & M7.5    & $>$M7.5 & $>$M7.5 & $>$M7.5 & $>$M7.5   \\
 2017\,10\,30   & $0.017 \pm 0.010$ & M3.5    & M4.5    & M4      & M3.5    & --      &  M3.5--M4 \\
\hline
\multicolumn{8}{p{12.5cm}}{$^a$ Pulsation phase according to ephemeris: JD$_{\rm{max}}$= 2457810.0 + 242.53 $\times$ E}
}

\noindent The method analogous to that applied by M\"urset \& Schmid
(1999) was used to evaluate the spectral type in two approaches:\\
({\sl{i}}) the strengths of TiO features in the spectra of Hen\,3-160 were
compared with those in the M giant standard spectra (Figure\,4 -- left). 
For each observed depth of TiO band the most suitable spectral type was
assigned (see Table\,3).\\
({\sl{ii}}) by dividing the spectrum of Hen\,3-160 with the standard
spectrum and assessing the smoothest ratio (Figure\,4 -- right).

\begin{figure}[htb]
\includegraphics{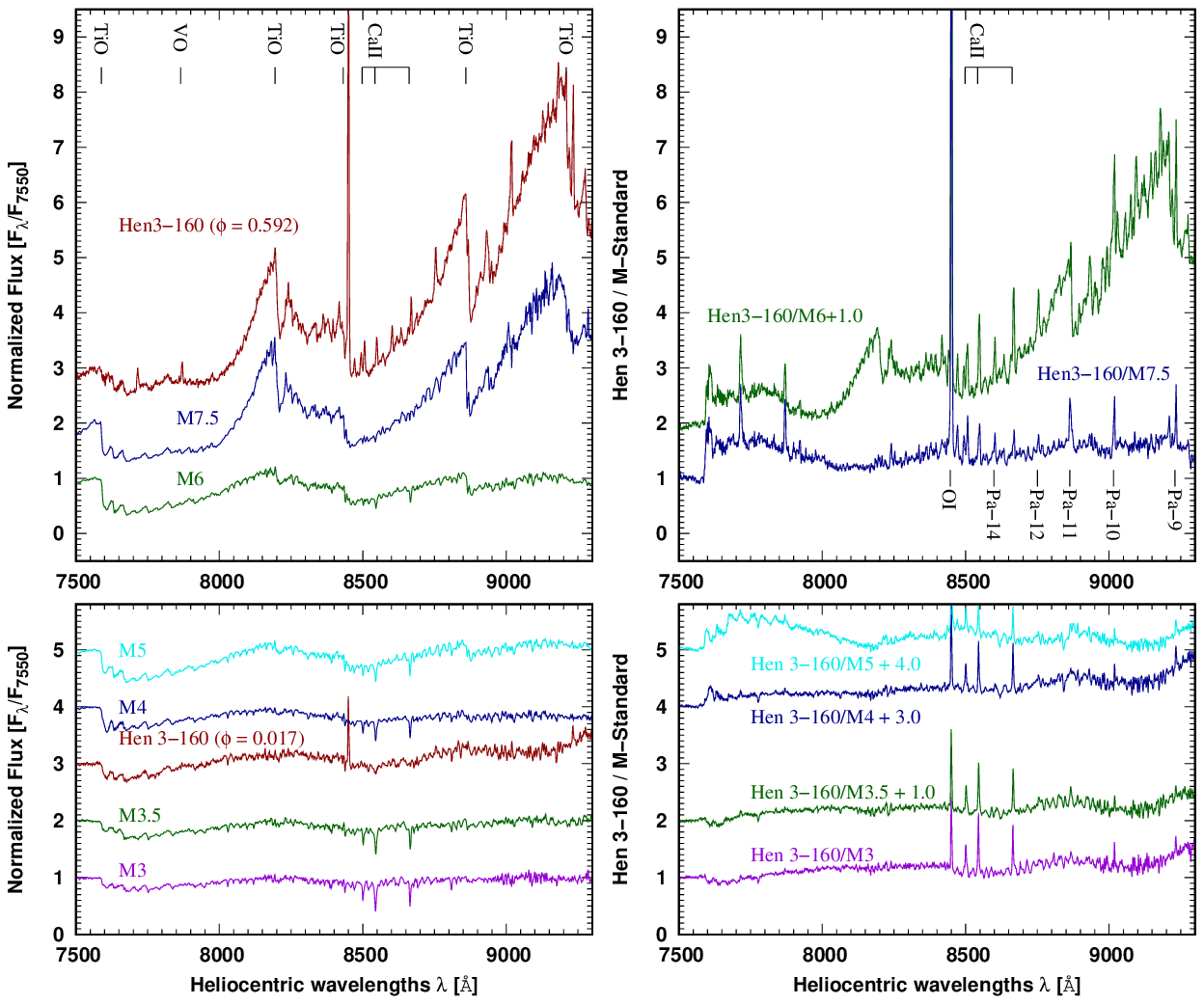}
\FigCap{{\sl{Left}}: the spectra of Hen\,3-160 (red) are compared with the
spectra of M-type giant spectroscopic standards (HR\,4902, HR\,6832,
HR\,8582, HR\,587, HR\,7625, and SW\,Vir) that represent spectral types from
M3 to M7.5.\\
{\sl{Right-top}}: ratios of the Hen\,3-160 spectrum obtained on March 2016
around pulsation minimum with the spectra of standards of types M6 and
M7.5 are presented.  The nebular contribution is manifested by line emission
from Ca\,II triplet, O\,I\,$\lambda$\,8442.36 and numerous H\,I Paschen
series lines.\\
{\sl{Right-bottom}}: ratios of the Hen\,3-160 spectrum obtained on October
2017 with the spectra of standards from type M3 up to M5.\\
The spectra are shifted by 1.0 for clarity.}
\end{figure}

For the spectrum from March 2016, the depth of TiO band at 7589\,\AA\,
indicates the spectral type later than M4.5 while the depths of TiO bands at
longer wavelengths indicate type M7.5 or later (Table\,3).  This spectrum
was placed close to pulsation minimum, where the contribution from a
continuum of the hot component must be significant and all these bands must
be shallowed, particularly at the blue edge.  The division by spectral
standards of various types gives the best result for type M7.5, however, it
can be seen from Figure\,4 (right-top) that a better result should be
obtained for the later type: $>$M7.5.

For the spectrum from October 2017, the spectral types from the depths of
TiO band heads are in the range M3.5--M4.5 (Table\,3).  The division of this
spectrum by the spectral standards (see Figure\,4 -- right-bottom) indicates
that the spectral type between M3.5--M4 should be the most suitable.  This
spectrum was obtained close to the pulsation maximum phase, i.e.  at the
condition when the depths of absorption features including TiO bands were at
the least degree influenced by any additional companion light, and it should
best pin down Mira's spectral type.  In addition, the temperature of the
pulsating giant reaches the maximum value soon before the photometric
maximum thus the spectral type derived here represents the lower limit,
i.e., it corresponds to the highest expected value of temperature for this
star.

\MakeTable{cccccc}{12.5cm}{Spectral type classification performed so far for Hen\,3-160. }
{\hline
Date          & JD/HJD      & Phase$^a$         & Spectral Region         & Spectral Type & Ref       \\
yyyy mm dd    & $-2400000$  &                   &                         &               &           \\
\hline
 ?            & ?           & ?                 & 2\,$\mu$m               & M7            & $^{[1]}$  \\
 1992 03 14   & 48696       & $0.42 \pm 0.04$   & 6900 -- 10700\,\AA      & M7            & $^{[2]}$  \\
 1994 04 30   & 49473       & $0.62 \pm 0.04$   & 6300 -- 10300\,\AA      & M7            & $^{[2]}$  \\
 1994 05 6-10 & 49479-49483 & $0.65$ -- $0.67$  & 6600 -- 7400\,\AA       & M4$^b$        & $^{[3]}$  \\
 2016 03 20   & 57468.4597  & $0.592 \pm 0.011$ & 7150 -- 9700\,\AA       & $\gtrsim$M8   & This work \\
 2017 10 30   & 58056.6118  & $0.017 \pm 0.010$ & $\sim$7100 -- 9700\,\AA & M3.5--M4      & This work \\
\hline
\multicolumn{6}{p{12.5cm}}{$^a$ Pulsation phase according to ephemeris: JD$_{\rm{max}}$= 2457810.0 + 242.53 $\times$ E.}\\
\multicolumn{6}{p{12.5cm}}{$^b$ This classification has to be mistaken as it was based on the optical spectrum obtained in pulsation phase close to brightness minimum when the absorption features were strongly veiled by the nebular component, and thus imitating earlier spectral type.}\\
\multicolumn{6}{p{12.5cm}}{{\bf{References:}} $^{[1]}$Allen (1980), $^{[2]}$M\"urset \& Schmid (1999), $^{[3]}$Harries \& Howarth (1996).}
}

%\newpage

Concluding, the {\bf{giant in Hen\,3-160 changes its spectral type with the
pulsation phase from about M3.5 during the brightness maximum up to M8 (or
later) during the minimum}}.  In Table\,4 the values obtained from our spectra
are compared with previous results of spectral type classification taken
from the literature based on the spectra obtained at various phases.  Using the
spectral type -- effective temperature calibrations (Richichi et al.  1999;
van Belle et al.  1999) we obtain the range of temperature changes from
$\sim$3500 to $\lesssim$ 3000\,K.  Colors derived from DENIS
(${\overline{J-K}}_0 \sim 1.16$) and 2MASS ($J-K_0 \sim 1.33$) measurements
also obtained at phases close to pulsation extrema (DENIS: $\phi = 0.934$ and
$0.105$; 2MASS: $\phi = 0.502$) result in temperatures $\sim$3550 and
$\sim$3200 K respectively.

\section{Chemical properties of the atmosphere}

The strong nebular continuum present in most cases in the optical spectra of
SySt causes a decrease in the molecular bands depth, imitating earlier spectral
types and lower abundances.  This usually causes severe difficulty in
studying parameters and measuring abundances of symbiotic giants based on
optical spectra.  The example can be the spectrum of Hen\,3-160 taken on
March 2016 close to pulsation minimum of Mira (Figure\,5), which was
dominated with radiation from nebula and absorption features completely
disappeared.  However, in two cases of spectra acquired close to pulsation
maximum, the influence by the nebula is already much smaller and absorption
features are pronounced.  In Figure\,5 these spectra (obtained on October
2017 and November 2005) are compared with the spectrum of well known
symbiotic S star CD$-27^{\circ}8661$ and the normal giant of M3.5 spectral
type.  The ZrO bands (eg.  with the heads at 6342, 6378, 6473, 6495, and
6505\,\AA) are very clear with strength comparable to those present in the
spectrum of S star.  In addition, the shape of strong TiO band at
$\sim$6148\,AA\, is changed in characteristic the s-process enhanced stars
way: its head is modified by overlapping YO and ZrO bands.  According to the
well known spectral classification scheme for cool giants we can ascribe MS
spectral type to this object as we see relatively strong ZrO bands in
addition to strong TiO bands in its spectrum (compare Hen\,3-160 spectra
from 2017 and 2005 in Figures 4 and 5 with figure\,1 from Yao et al.  2017).

\begin{figure}[htb]
\includegraphics{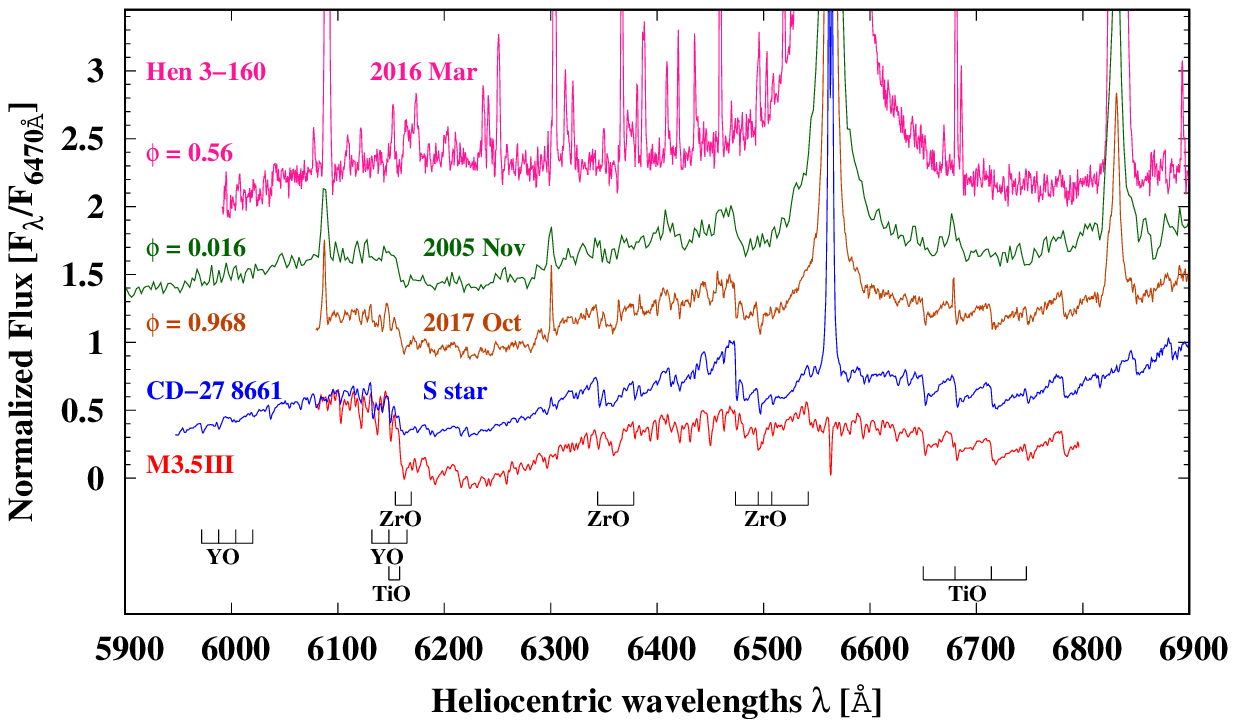}
\FigCap{Comparison of spectra of Hen\,3-160 obtained at three various phases
-- one close to pulsation minimum and two close to pulsation maximum -- with
the spectrum of well known symbiotic S star CD$-27^{\circ}8661$ and a red
giant of M3.5 spectral type.}
\end{figure}

The above confrontation with the spectrum of S star leaves no doubts that
the giant in Hen\,3-160 is enhanced in s-process elements.  As the cool
component in Hen\,3-160 is a Mira star any quantitative analysis would
be problematic because the appropriate atmosphere models do not exist. 
Nevertheless, we can use our optical spectra with well visible ZrO bands for
the analysis of the spectral indices.  We performed analysis for a large
sample of spectra collected for above 50 symbiotic giants and a dozen
single red giants.  The method was similar to that used by Van Eck et al. 
(2017) in their study of the properties and atmospheres of S stars.  We also
used their sample of S stars.  The band-strength indices were calculated
according to the equation:

$B_{\rm{X}} = 1 - \frac{\int_{\lambda_{B_{\rm{i}}}}^{\lambda_{B_{\rm{f}}}} F_{\lambda} d\lambda}{(\lambda_{B_{\rm{f}}} - \lambda_{B_{\rm{i}}})} \frac{\lambda_{C_{\rm{f}}} - \lambda_{C_{\rm{i}}}}{\int_{\lambda_{C_{\rm{i}}}}^{\lambda_{C_{\rm{f}}}} F_{\lambda} d\lambda}$,

\noindent where '$F_{\lambda}$' is the observed flux in the range of
wavelength ($\lambda$,$\lambda + d\lambda$), and indexes '${\rm{i}}$' and
'${\rm{f}}$' denote the wavelengths of beginnings and ends, respectively, of
the band '$B$' and the continuum '$C$' windows listed in Table\,5.

\begin{figure}[htb]
\includegraphics{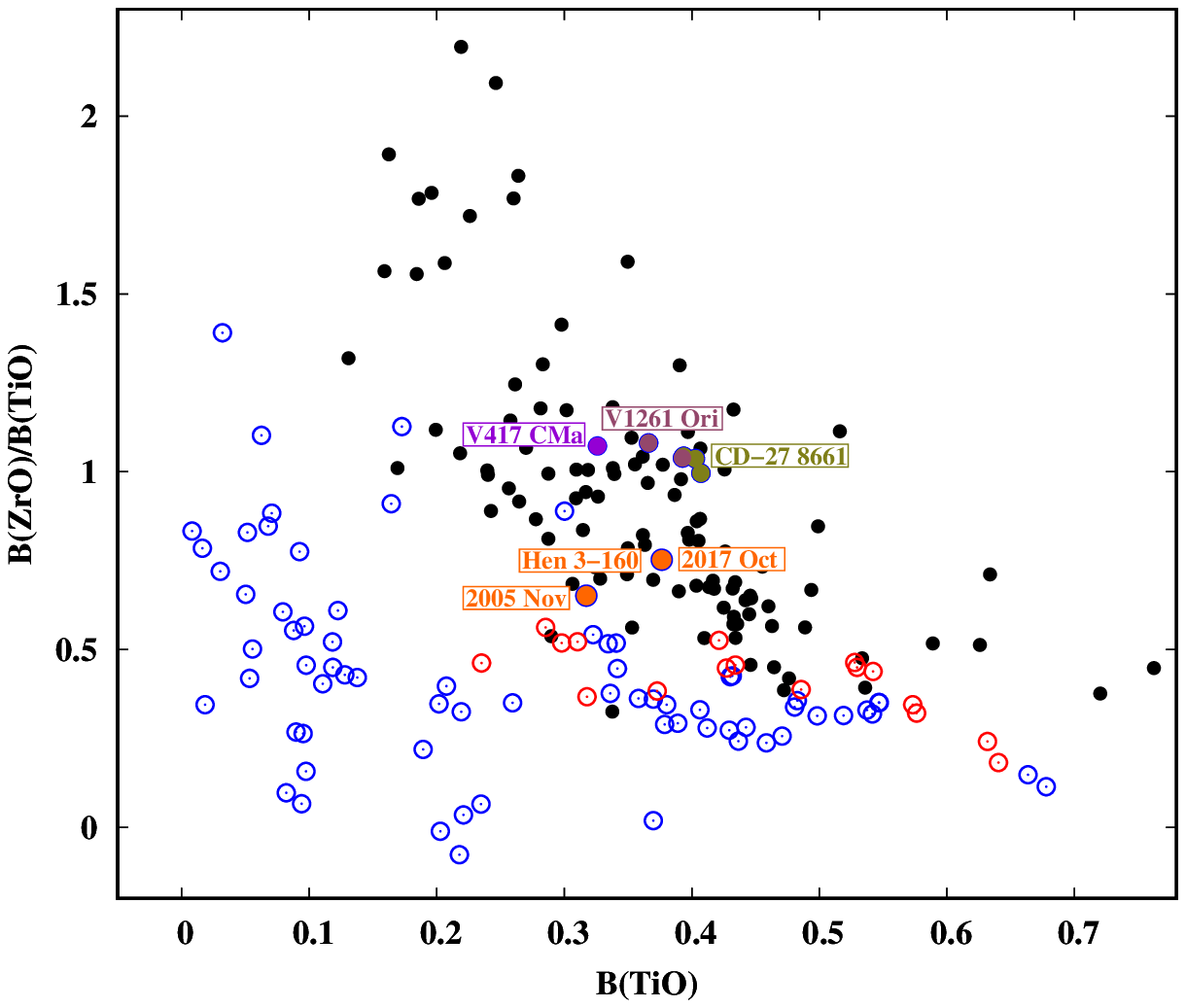}
\FigCap{The plane of indices $B_{\rm{ZrO}}/B_{\rm{TiO}}$ vs.  $B_{\rm{TiO}}$
with our program SySt (blue circles) and normal M giants (red circles)
compared to S stars (black dots).  With various coloued circles are
distinguished the more interesting cases: well known SySt containing S stars
(V1261 Ori, CD$-27^{\circ}8661$), suspected SySt with S-type giant
V417\,CMa, and the positions of Hen\,3-160 are featured with orange
circles.}
\end{figure}

\MakeTable{lll|lll}{12.5cm}{Bands borders that have been adopted to calculate
the spectral indices $B_{\rm{X}}$ according to the above formula.}
{\hline
$\lambda_{\rm{i}}$ & $\lambda_{\rm{f}}$ & Band & $\lambda_{\rm{i}}$ & $\lambda_{\rm{f}}$ & Band \\
\hline
6144.0             & 6147.5             & Cont & 6464.0             & 6472.0             & Cont \\
6148.5             & 6151.5             & TiO  & 6344.4             & 6346.0             & ZrO  \\
6158.2             & 6167.8             & TiO  & 6378.0             & 6379.9             & ZrO  \\
6173.7             & 6181.3             & TiO  & 6473.5             & 6476.7             & ZrO  \\
6185.4             & 6198.7             & TiO  & 6496.1             & 6497.4             & ZrO  \\
                   &                    &      & 6507.6             & 6510.2             & ZrO  \\
}

The $B_{\rm{ZrO}}/B_{\rm{TiO}}$ vs.  $B_{\rm{TiO}}$ plane is shown in
Figure\,6.  The indices calculated from the spectrum of Hen\,3-160 obtained
on October 2017 (at $\phi = 0.984$) result in a position (orange circle)
well in the region of S stars, and close to positions of known SySt
containing S stars (V1261 Ori, CD$-27^{\circ}8661$) and suspected SySt with
S-type giant V417\,CMa.  This is a strong indication that the {\bf{giant in
Hen 3-160 is s-process enhanced star}}.

\section{Spectral energy distribution (SED) analysis}

The SED was constructed using all available photometry (Appenxid: Table\,7). 
The data was de-reddened using the color excess value E$_{B-V} = 0.35$ and
by adopting the mean interstellar extinction curve for R=3.1 (Fitzpatrick
2004).

\begin{figure}[htb]
\includegraphics{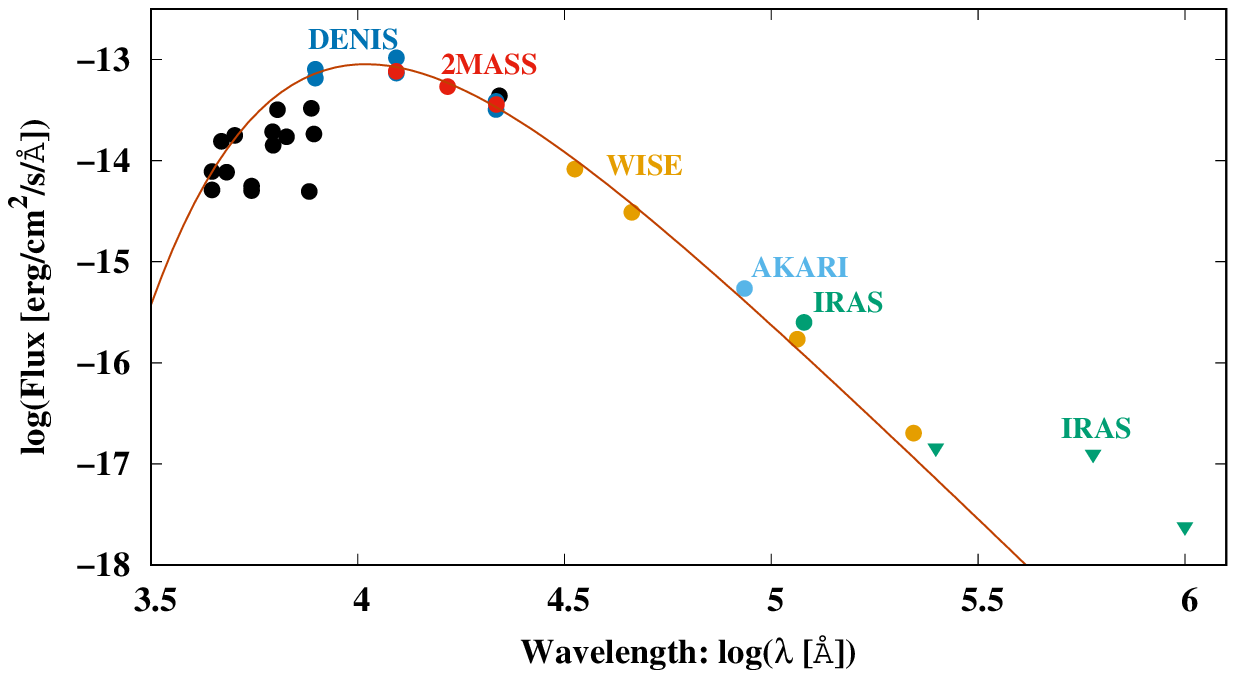}
\FigCap{Spectral energy distribution of Hen\,3-160 obtained from all
available photometry (Appendix: Table\,7).  The line shows the fit of
Black-Body (T $= 2770 \pm 330$\,K) to the data of the near-$IR$ region. 
Triangles denote the upper estimates.}
\end{figure}

\noindent The magnitudes were transformed into fluxes F$_{\lambda}$
expressed in units of $[erg$\,$cm^{-2} s^{-1}$ \AA$^{-1}]$ using the Bessel
et al.  (1998) calibration.  Conversion from magnitudes into fluxes F$_J$
(expressed in Jy) in the case of WISE photometry was made with a use of zero
magnitude flux densities according to Jarrett et al.  (2011).  Jansky's
F$_J$ were finally recalculated into F$_{\lambda}$ fluxes according to the
expression: F$_{\lambda}$ = F$_J$ $\lambda^{-2} 33356.4095^{-1}$.  The
resulted SED is shown in Figure\,7 with a fit of Black-Body of temperature T
$= 2770 \pm 330$\,K to the data of the near-$IR$ region.  {\bf{The infrared
excess is practically absent which indicates the lack of developed
circumstellar dust shell}} and confirms the classification of Hen\,3-160 as
an S-type SySt.  This is more or less consistent with theoretical studies by
Vassiliadis \& Wood (1993) who found out that for Miras with pulsation
period below $\sim$ 600 -- 800 days the mass-loss rate increases
exponentially, while above these values it is essentially constant at the
radiation-pressure-driven limit, and at extremely long periods the stars are
permanently obscured by circumstellar dust.  Our Mira with relatively short
P$_{\rm{pul}}$=242.5\,d should not be significantly obscured by dust in
agreement with the lack of infrared excess observed in the SED.
The confrontation with IRAS photometry of SySt (Kenyon
et al.  1988) shows that Hen\,3-160 has very low infrared fluxes with lack
of IR excess -- like S-type SySt (which have very low infrared fluxes and in
majority only the upper limits are estimated) -- and on the contrary to
D-type SySt with Mira variables as the cool companion that show significant infrared excesses.

\begin{figure}[htb]
\includegraphics{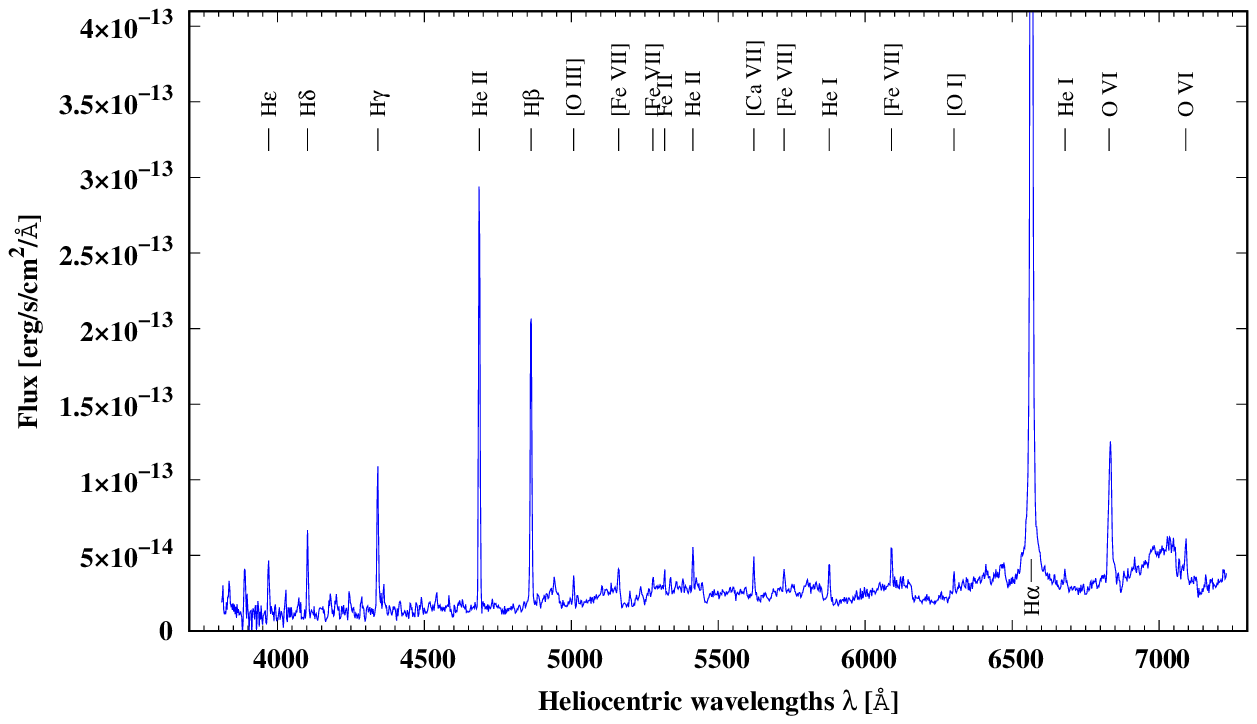}
\FigCap{The flux calibrated, rescaled with $V$ = 13.9 mag and de-reddened the
optical spectrum (obtained on November 2005) with identified emission lines
marked with ticks.}
\end{figure}

\section{The nature of the hot companion}

We have observed a number of emission lines in our optical spectra which we
use here to shed more light on the nature of the hot component in Hen\,3-160
by estimating its temperature and luminosity.  We analyzed two spectra
obtained on November 20, 2005, and March 16, 2016.  We have applied an
absolute flux scale to the first spectrum using the $V=13.9$\,mag
corresponding to the date of its observation ($V$ light curve --
Sections\,2\,\&\,3) such that convolution of the spectrum with the Johnson
$V$ filter agrees with the $V$ mag.  The spectrum was de-reddened adequately
to E$_{B-V} = 0.35$, and the emission line fluxes were measured using
{\sl{splot}} tool of the {\sl{IRAF}} packages.  The identified emission
lines are shown in Figure\,8 and together with measured flux values are
listed in Table\,6.  The second spectrum was obtained in higher resolution
at Mira's minimum brightness and it was possible to identify a slightly
larger number of emission lines.  However, it covered too narrow wavelength
range to enable the absolute flux calibration using available photometry,
and because it does not show the presence of lines corresponding to higher
ionization potential it was omitted in the further analysis.

\MakeTable{lccc}{12.5cm}{Emission line fluxes measured from the spectrum
obtained on November 2005 which was de-reddened and scaled to $V =
13.9$\,mag.  Ionization potentials are also shown on the right.  The
errors in the measured fluxes come mainly from uncertainty in the flux
calibration of the spectrum which results from scaling with $V$-band
photometry and are of order $20\%$.
}
{\hline
Line                         & Measured wavelength & Flux                               & $IP$ \\
                             & $\lambda$ [\AA]     & $[10^{-14} erg$\,$s^{-1} cm^{-2}]$ & [eV] \\
\hline
H$\epsilon$                  & 3969.9    &   24    &        \\
H$\delta$                    & 4102.4    &   33    &        \\
H$\gamma$                    & 4341.3    &   66    &        \\
He\,II\,$\lambda$4686        & 4686.9    &  173    &    54  \\
H$\beta$                     & 4862.9    &  138    &        \\
$[$O\,III$]$\,$\lambda$5007  & 5008.0    &   11    &    55  \\
$[$Fe\,VII$]$\,$\lambda$5158 & 5161.1    &    9.9  &    99  \\
$[F$e\,VII$]$\,$\lambda$5276 & 5277.5    &    3.9  &    99  \\
Fe\,II\,$\lambda$5317        & 5317.4    &    5.0  &    16  \\
He\,II\,$\lambda$5412        & 5413.8    &   15    &    54  \\
$[$Ca\,VII$]$\,$\lambda$5619 & 5621.0    &   14.5  &   109  \\
$[$Fe\,VII$]$\,$\lambda$5721 & 5723.9    &    8.3  &    99  \\
He\,I\,$\lambda$5876         & 5877.6    &   19    &    25  \\
$[$Fe\,VII$]$\,$\lambda$6086 & 6089.8    &   17    &    99  \\
$[$O\,I$]$\,$\lambda$6300    & 6302.4    &    7.9  &    14  \\
H$\alpha$                    & 6565.4    & 1249    &        \\
He\,I\,$\lambda$6678         & 6680.1    &    8.0  &    25  \\
O\,VI\,$\lambda$6825         & 6825.1    &   10    &   114  \\
O\,VI\,$\lambda$6825         & 6834.0    &  105    &   114  \\
O\,VI\,$\lambda$7088         & 7091.0    &   18.5  &   114  \\
\hline
}

We used the relation $T \sim 10^{3} \times IP$\,(eV) proposed by M{\"u}rset
\& Nussbaumer (1994) to estimate the lower limit of temperature.  The method
is valid for $T < 150$\,kK.  The maximum ionization potential $IP = 114$\,eV
is provided by O\,VI lines what corresponds to value $T \gtrsim 114$\,kK. 
The upper limit can be estimated by using the Iijima (1981) method based on
the ratios of emission line fluxes of H$\beta$, He\,II\,$\lambda$4686 and
He\,I\,$\lambda$5876.  This method is valid for effective temperatures
between 70 and 200\,kK.  The line flux for the line He\,I\,$\lambda$4471
which originally occurs in the Iijima (1981) equation was replaced here with
the flux for the significantly stronger He\,I $\lambda$5876 line taking into
account the difference in intensity by a factor of $\sim$2.7 (see e.g. 
Osterbrock 1989).  The resulted value of upper limit for WD temperature, in
this case, is $T \lesssim 200$\,kK.  To estimate luminosity of hot component
we used several empirical formulas provided in the literature.  The equation
8 of Kenyon et al.  (1991) applicable to fluxes from mentioned above three
emission lines gives the value of luminosity $L \sim 770$\,L$\odot$.  Fluxes
from He\,II\,$\lambda$4686 and H$\beta$ lines could be used to estimate
luminosity via equations 6 and 7, respectively, of Miko{\l}ajewska et al. 
(1997).  We derived the values $L$(He\,II\,$\lambda$4686)
$\sim$1280\,L$\odot$ and $L$(H$\beta$) $\sim$1020\,L$\odot$ for the maximum
limit of WD temperature $T = 200$\,kK.  All the above estimates adopt the
distance to Hen\,3-160 d=9.4\,kpc (Section\,4), assume a blackbody spectrum
for the hot component and Case B recombination for the emission lines, and
have an uncertainty by a factor of two.  The comparison with the largest so
far studied sample of symbiotic hot components (Miko{\l}ajewska et al. 
1997) locates Hen\,3-160 among the hotest and moderately luminous systems.

\section{Conclusions}

We performed the first comprehensive analysis of the Hen\,3-160 symbiotic
binary based on new photometric and spectroscopic data in optical and
infrared domains.  The large-amplitude periodic variations observed in the
optical $V$ and $I_{\rm{C}}$-band light curves with $P_{\rm{pul}}$ = 242.5
days, correlated with changes in other bands as well in the spectra,
indicate that the cool component is a Mira star.  The changes in the
spectral type in the range $\sim$M3.5--$\gtrsim$M8 estimated based on the
TiO bands in optical spectra are correlated with the Mira pulsations.

From the period-luminosity relation, the distance to Hen 3-160 is of about
9.4 kpc.  The Galactic coordinates place it $\sim$1.3 kpc above the disc and
combined with relatively high proper motions indicates that Hen\,3-160 has
to be a Galactic extended thick disc object.

Our optical spectra show the presence of ZrO and YO molecular bands that are
relatively strong compared to TiO bands.  The presence of comparably strong
ZrO and TiO bands is consistent with the MS spectral type for this object. 
Analysis of the spectral indices constructed using ZrO and TiO bands has
placed Hen\,3-160 among the S stars proving that it is enhanced in the
s-process elements.  To our knowledge, Hen 3-160 is the first known
symbiotic Mira that is simultaneously the s-process enhanced star of MS
spectral type.

The Mira in Hen\,3-160 with relatively short pulsation period should not be
significantly obscured by dust which is consistent with very low infrared
excess observed in the SED.

Based on analysis of emission lines in optical spectra the temperature of
the white dwarf is in the range of $\sim 114$ -- $200$\,kK and its
luminosity is between $\sim$770--1280\,L$\odot$.\\

\Acknow{CG has been financed by the Polish National Science Centre (NCN)
grant SONATA No.  DEC-2015/19/D/ST9/02974.  KI has been financed by the
Polish Ministry of Science and Higher Education Diamond Grant Programme via
grant 0136/DIA/2014/43 and by the Foundation for Polish Science (FNP) within
the START program.  This study has been supported in part by NCN grant OPUS
2017/27/B/ST9/01940.  MG is supported by the Polish National Science Centre
grant OPUS 2015/17/B/ST9/03167.  This study uses spectroscopic observations
collected with 1.9 m telescope at the South African Astronomical
Observatory, which was possible due to Polish participation in SALT funded
by grant No.  MNiSW DIR/WK/2016/07.  This research has made use of the NASA/
IPAC Infrared Science Archive, which is operated by the Jet Propulsion
Laboratory, California Institute of Technology, under contract with the
National Aeronautics and Space Administration.  We would like cordially
thank Sophie Van Eck and Alain Jorissen from Universit\'e Libre de
Bruxelles for providing a rich collection of S stars spectra used in our
analysis of the spectral indices.}

\newpage

\begin{appendix}
\begin{center}
\bf{Appendix: complementary tables}
\end{center}

\MakeTable{@{}c@{\hskip 2mm}c@{\hskip 2mm}c@{\hskip 2mm}c@{\hskip 2mm}c@{\hskip 2mm}c@{\hskip 2mm}c@{\hskip 2mm}c@{}}{12.5cm}{Photometry taken from the literature and the catalogs, and fluxes calculated as described in section\,7.}
{\hline
Band     & Wav.               & JD         & Phase$^a$ & mag          & Flux                  & Flux                                   & Source\\
         & $\bar\lambda$[\AA] & $-2450000$ &           &              & $[Jy]$                & $[erg$\,$cm^{-2} s^{-1}$ \AA$^{-1}]$   &       \\
\hline
%          Band      & Wavelength& JD      & phase & [mag]            & sigma                 & Flux                    & Source \\
         Johnson:$B$ &   4442    & --      & --    & 16.144$\pm$0.031 &                       & 2.13e-15$\pm$6.09e-17 & APASS DR9$^{[1]}$\\ %AAVSO Photometric All Sky Survey (APASS) DR9 (Henden+, 2016)   (II/336/apass9)
         Johnson:$B$ &   4442    & --      & --    & 16.60            &                       & 1.40e-15              & SPM 4.0$^{[2]}$  \\ %Yale/San Juan Southern Proper Motion Catalog 4 (Girard+, 2011) (I/320/spm4)
         POSS-II:$J$ &   4680    & --      & --    & 15.18$\pm$0.40   &                       & 4.57e-15$\pm$1.72e-15 & GSC V.2.3$^{[3]}$\\ %Guide Star Catalog, V.2.3                                      (I/305/out)
            SDSS:$g$ &   4820    & --      & --    & 15.816$\pm$0.047 &                       & 2.36e-15$\pm$1.02e-16 & APASS DR9$^{[1]}$\\ %AAVSO Photometric All Sky Survey (APASS) DR9 (Henden+, 2016)   (II/336/apass9)
          Gaia:$Gbp$ &   5046    & --      & --    & 14.700$\pm$0.069 &                       & 5.86e-15$\pm$3.73e-16 & Gaia DR2$^{[4]}$ \\ %Gaia data release 2 (Gaia DR2)                                 (I/345/gaia2)
         Johnson:$V$ &   5537    & --      & --    & 15.640$\pm$0.034 &                       & 1.91e-15$\pm$5.99e-17 & APASS DR9$^{[1]}$\\ %AAVSO Photometric All Sky Survey (APASS) DR9 (Henden+, 2016)   (II/336/apass9)
         Johnson:$V$ &   5537    & --      & --    & 15.53            &                       & 2.12e-15              & SPM 4.0$^{[2]}$  \\ %Yale/San Juan Southern Proper Motion Catalog 4 (Girard+, 2011) (I/320/spm4)
            Gaia:$G$ &   6226    & --      & --    & 13.637$\pm$0.030 &                       & 8.53e-15$\pm$2.36e-16 & Gaia DR2$^{[4]}$ \\ %Gaia data release 2 (Gaia DR2)                                 (I/345/gaia2)
            SDSS:$r$ &   6247    & --      & --    & 13.954$\pm$0.094 &                       & 6.31e-15$\pm$5.47e-16 & APASS DR9$^{[1]}$\\ %AAVSO Photometric All Sky Survey (APASS) DR9 (Henden+, 2016)   (II/336/apass9)
         POSS-II:$F$ &   6400    & --      & --    & 12.96$\pm$0.44   &                       & 1.46e-14$\pm$6.09e-15 & GSC V.2.3$^{[3]}$\\ %Guide Star Catalog, V.2.3                                      (I/305/out)
            Gaia:$G$ &   6730    & --      & --    & 13.391$\pm$0.036 &                       & 8.38e-15$\pm$2.78e-16 & HSOY$^{[5]}$     \\ %Hot Stuff for One Year (HSOY) (Altmann+, 2017)                 (I/339/hsoy) 
            SDSS:$i$ &   7635    & --      & --    & 14.154$\pm$0.391 &                       & 2.76e-15$\pm$1.02e-15 & APASS DR9$^{[1]}$\\ %AAVSO Photometric All Sky Survey (APASS) DR9 (Henden+, 2016)   (II/336/apass9)
          Gaia:$Grp$ &   7725    & --      & --    & 12.041$\pm$0.135 &                       & 1.86e-14$\pm$2.32e-15 & Gaia DR2$^{[4]}$ \\ %Gaia data release 2 (Gaia DR2)                                 (I/345/gaia2)
         POSS-II:$i$ &   7837    & --      & --    & 12.61$\pm$0.43   &                       & 1.05e-14$\pm$4.27e-15 & GSC V.2.3$^{[3]}$\\ %Guide Star Catalog, V.2.3                                      (I/305/out)
                 $I$ &   7900    & 1245.60 & 0.934 & 11.191$\pm$0.03  &                       & 3.79e-14$\pm$1.05e-15 & DENIS            \\
                 $I$ &   7900    & 1529.79 & 0.105 & 10.974$\pm$0.02  &                       & 4.62e-14$\pm$8.51e-16 & DENIS            \\
                 $J$ &  12390    & 1140.83 & 0.502 &  9.299$\pm$0.023 &                       & 5.98e-14$\pm$1.27e-15 & 2MASS$^{[6]}$    \\%                                                                (II/328/allwise)
                 $J$ &  12400    & 1245.60 & 0.934 &  9.340$\pm$0.04  &                       & 5.74e-14$\pm$2.12e-15 & DENIS            \\
                 $J$ &  12400    & 1529.79 & 0.105 &  8.959$\pm$0.06  &                       & 8.16e-14$\pm$4.51e-15 & DENIS            \\
                 $H$ &  16495    & 1140.83 & 0.502 &  8.333$\pm$0.038 &                       & 4.66e-14$\pm$1.63e-15 & 2MASS$^{[6]}$    \\%                                                                (II/328/allwise)
                 $K$ &  21600    & 1245.60 & 0.934 &  7.918$\pm$0.07  &                       & 2.93e-14$\pm$1.89e-15 & DENIS            \\
                 $K$ &  21600    & 1529.79 & 0.105 &  7.712$\pm$0.06  &                       & 3.54e-14$\pm$1.96e-15 & DENIS            \\
         $K_{\rm S}$ &  21637    & 1140.83 & 0.502 &  7.801$\pm$0.021 &                       & 3.29e-14$\pm$6.37e-16 & 2MASS$^{[6]}$    \\%                                                                (II/328/allwise)
      $W1$:3.4$\mu$m &  33500    & --      & --    &  7.545$\pm$0.036 & 0.297$\pm$0.015$^b$   & 7.93e-15$\pm$4.01e-16 & AllWISE SC       \\%                                                                (II/328/allwise) (mean JD=2455454.9)
      $W2$:4.6$\mu$m &  46000    & --      & --    &  7.268$\pm$0.021 & 0.213$\pm$0.008$^b$   & 3.02e-15$\pm$1.13e-16 & AllWISE SC       \\%                                                                (II/328/allwise) (mean JD=2455454.9)
       $W3$:12$\mu$m & 115600    & --      & --    &  6.553$\pm$0.016 & 0.0758$\pm$0.0023$^b$ & 1.70e-16$\pm$5.16e-18 & AllWISE SC       \\%                                                                (II/328/allwise) (mean JD=2455341.1)
       $W4$:22$\mu$m & 220900    & --      & --    &  6.018$\pm$0.038 & 0.0327$\pm$0.0024$^b$ & 2.01e-17$\pm$1.47e-18 & AllWISE SC       \\%                                                                (II/328/allwise) (mean JD=2455341.1)
         2.2\,$\mu$m &  22000    & --      & --    &                  & 0.65($\sim$10$\%$)    & 4.03e-14$\pm$4.03e-15 & $^{[7]}$         \\
           9\,$\mu$m &  86100    & --      & --    &                  & 0.133$\pm$0.008       & 5.38e-16$\pm$3.24e-17 & AKARI PSC        \\%                                                                (II/297/irc)
          12\,$\mu$m & 120000    & --      & --    &                  &   0.12                & 2.50e-16              & IRAS$^{[7]}$     \\
          25\,$\mu$m & 250000    & --      & --    &                  &$<$0.03                & $<$1.44e-17           & IRAS$^{[7]}$     \\
          60\,$\mu$m & 600000    & --      & --    &                  &$<$0.15                & $<$1.25e-17           & IRAS$^{[7]}$     \\
         100\,$\mu$m &1000000    & --      & --    &                  &$<$0.08                & $<$2.40e-18           & IRAS$^{[7]}$     \\
\hline
\multicolumn{8}{p{14.5cm}}{$^a$Pulsation phase according to ephemeris: JD$_{\rm{max}}$= 2457810.0 + 242.53 $\times$ E}\\ 
\multicolumn{8}{p{14.5cm}}{$^b$Calculated with the use of zero magnitude flux densities according to Jarrett et al.(2011)}\\
\multicolumn{8}{p{14.5cm}}{{\bf{References:}} $^{[1]}$Henden et al.\,(2015), $^{[2]}$Girard et al.\,(2011), $^{[3]}$Lasker et al.\,(2008), $^{[4]}$Gaia Collaboration\,(2018), $^{[5]}$Altmann et al.\,(2017), $^{[6]}$Phillips\,(2007), $^{[7]}$Kenyon et al.\,(1988)}
}

\MakeTable{llcccc}{12.5cm}{WISE photometry from AllWISE Multiepoch Photometry Table.}
{\hline
MJD-2450000 & phase$^a$ & $W1 [$mag$]$ & $W2 [$mag$]$ & $W3 [$mag$]$ & $W4 [$mag$]$\\
\hline
%          mjd| w1mpro_ep| w1sigmpro_ep| w1o_ep| w2sigmpro_ep| w2o_ep| w3sigmpro_ep| w3o_ep| w4sigmpro_ep|qi_fact|
%       double|    double|       double|   uble|       double|   uble|       double|   uble|       double| double|
%          day|       mag|          mag|    mag|          mag|    mag|          mag|    mag|          mag|       |
5340.497 &  0.818 &  7.446 $\pm$ 0.028 & 7.124 $\pm$ 0.026 & 6.545 $\pm$ 0.018 & 6.113 $\pm$ 0.126 \\ %1.0  
5340.761 &  0.819 &  7.426 $\pm$ 0.021 & 7.107 $\pm$ 0.023 & 6.545 $\pm$ 0.021 & 5.800 $\pm$ 0.077 \\ %1.0  
5340.761 &  0.819 &  7.464 $\pm$ 0.024 & 7.161 $\pm$ 0.023 & 6.557 $\pm$ 0.018 & 5.962 $\pm$ 0.110 \\ %1.0  
5340.893 &  0.820 &  7.426 $\pm$ 0.026 & 7.118 $\pm$ 0.020 & 6.541 $\pm$ 0.021 & 6.076 $\pm$ 0.135 \\ %1.0  
5341.026 &  0.820 &  7.432 $\pm$ 0.025 & 7.138 $\pm$ 0.020 & 6.535 $\pm$ 0.019 & 6.052 $\pm$ 0.096 \\ %1.0  
5341.092 &  0.821 &  7.474 $\pm$ 0.021 & 7.110 $\pm$ 0.024 & 6.546 $\pm$ 0.016 & 6.086 $\pm$ 0.107 \\ %1.0  
5341.158 &  0.821 &  7.454 $\pm$ 0.025 & 7.118 $\pm$ 0.019 & 6.545 $\pm$ 0.022 & 5.875 $\pm$ 0.086 \\ %1.0  
5341.224 &  0.821 &  7.494 $\pm$ 0.027 & 7.146 $\pm$ 0.027 & 6.553 $\pm$ 0.020 & 6.027 $\pm$ 0.095 \\ %1.0  
5341.290 &  0.822 &  7.437 $\pm$ 0.023 & 7.134 $\pm$ 0.018 & 6.553 $\pm$ 0.019 & 6.026 $\pm$ 0.096 \\ %1.0  
5341.356 &  0.822 &  7.427 $\pm$ 0.021 & 7.150 $\pm$ 0.020 & 6.541 $\pm$ 0.017 & 5.858 $\pm$ 0.134 \\ %1.0  
5341.423 &  0.822 &  7.439 $\pm$ 0.023 & 7.071 $\pm$ 0.016 & 6.555 $\pm$ 0.019 & 5.928 $\pm$ 0.097 \\ %1.0  
5341.489 &  0.822 &  7.454 $\pm$ 0.024 & 7.144 $\pm$ 0.019 & 6.541 $\pm$ 0.019 & 5.943 $\pm$ 0.089 \\ %1.0  
5341.753 &  0.823 &  7.492 $\pm$ 0.023 & 7.083 $\pm$ 0.019 & 6.533 $\pm$ 0.016 & 5.943 $\pm$ 0.107 \\ %1.0  
5341.819 &  0.824 &  7.429 $\pm$ 0.020 & 7.120 $\pm$ 0.018 & 6.574 $\pm$ 0.019 & 6.046 $\pm$ 0.096 \\ %1.0  
5341.886 &  0.824 &  7.437 $\pm$ 0.022 & 7.151 $\pm$ 0.019 & 6.548 $\pm$ 0.016 & 5.943 $\pm$ 0.101 \\ %1.0  
5342.018 &  0.825 &  7.423 $\pm$ 0.022 & 7.114 $\pm$ 0.016 & 6.541 $\pm$ 0.018 & 6.134 $\pm$ 0.104 \\ %1.0  
5342.084 &  0.825 &  7.427 $\pm$ 0.021 & 7.100 $\pm$ 0.023 & 6.594 $\pm$ 0.021 & 6.123 $\pm$ 0.100 \\ %1.0  
5342.150 &  0.825 &  7.427 $\pm$ 0.023 & 7.095 $\pm$ 0.021 & 6.560 $\pm$ 0.019 & 6.080 $\pm$ 0.096 \\ %1.0  
5342.216 &  0.825 &  7.444 $\pm$ 0.019 & 7.121 $\pm$ 0.018 & 6.577 $\pm$ 0.020 & 6.202 $\pm$ 0.116 \\ %1.0  
5342.283 &  0.826 &  7.424 $\pm$ 0.023 & 7.137 $\pm$ 0.022 & 6.561 $\pm$ 0.019 & 6.190 $\pm$ 0.110 \\ %1.0  
5342.415 &  0.826 &  7.438 $\pm$ 0.027 & 7.159 $\pm$ 0.023 & 6.589 $\pm$ 0.023 & 6.017 $\pm$ 0.101 \\ %1.0  
5342.547 &  0.827 &  7.436 $\pm$ 0.020 & 7.085 $\pm$ 0.018 & 6.555 $\pm$ 0.021 & 5.972 $\pm$ 0.100 \\ %1.0  
5528.462 &  0.593 &  7.597 $\pm$ 0.045 & 7.392 $\pm$ 0.026 & --                & --                \\ %0.5  
5528.594 &  0.594 &  7.620 $\pm$ 0.031 & 7.341 $\pm$ 0.020 & --                & --                \\ %1.0  
5528.727 &  0.594 &  7.682 $\pm$ 0.041 & 7.359 $\pm$ 0.022 & --                & --                \\ %1.0  
5528.859 &  0.595 &  7.750 $\pm$ 0.043 & 7.344 $\pm$ 0.023 & --                & --                \\ %1.0  
5528.859 &  0.595 &  7.705 $\pm$ 0.042 & 7.337 $\pm$ 0.026 & --                & --                \\ %1.0  
5529.057 &  0.596 &  7.702 $\pm$ 0.031 & 7.305 $\pm$ 0.017 & --                & --                \\ %1.0  
5529.124 &  0.596 &  7.687 $\pm$ 0.031 & 7.368 $\pm$ 0.027 & --                & --                \\ %1.0  
5530.975 &  0.596 &  7.557 $\pm$ 0.035 & 7.356 $\pm$ 0.024 & --                & --                \\ %1.0  
5531.240 &  0.597 &  7.637 $\pm$ 0.030 & 7.366 $\pm$ 0.028 & --                & --                \\ %1.0  
5531.306 &  0.597 &  7.799 $\pm$ 0.055 & 7.415 $\pm$ 0.024 & --                & --                \\ %1.0  
5531.372 &  0.597 &  7.620 $\pm$ 0.031 & 7.350 $\pm$ 0.022 & --                & --                \\ %1.0  
5531.438 &  0.606 &  7.629 $\pm$ 0.026 & 7.342 $\pm$ 0.021 & --                & --                \\ %1.0  
5531.504 &  0.606 &  7.648 $\pm$ 0.037 & 7.339 $\pm$ 0.018 & --                & --                \\ %1.0  
5531.570 &  0.606 &  7.624 $\pm$ 0.037 & 7.348 $\pm$ 0.020 & --                & --                \\ %1.0  
5531.637 &  0.606 &  7.642 $\pm$ 0.044 & 7.331 $\pm$ 0.023 & --                & --                \\ %1.0  
5531.703 &  0.607 &  7.717 $\pm$ 0.036 & 7.348 $\pm$ 0.027 & --                & --                \\ %0.5  
5531.769 &  0.607 &  7.669 $\pm$ 0.032 & 7.394 $\pm$ 0.028 & --                & --                \\ %1.0  
5531.835 &  0.607 &  7.705 $\pm$ 0.040 & 7.332 $\pm$ 0.021 & --                & --                \\ %1.0  
5531.901 &  0.607 &  7.638 $\pm$ 0.038 & 7.424 $\pm$ 0.022 & --                & --                \\ %1.0  
5531.967 &  0.608 &  7.609 $\pm$ 0.032 & 7.404 $\pm$ 0.022 & --                & --                \\ %1.0  
5532.033 &  0.608 &  7.753 $\pm$ 0.040 & 7.275 $\pm$ 0.021 & --                & --                \\ %1.0  
5532.100 &  0.608 &  7.671 $\pm$ 0.035 & 7.423 $\pm$ 0.023 & --                & --                \\ %1.0  
5532.166 &  0.609 &  7.724 $\pm$ 0.028 & 7.321 $\pm$ 0.020 & --                & --                \\ %1.0  
5532.232 &  0.609 &  7.737 $\pm$ 0.039 & 7.318 $\pm$ 0.021 & --                & --                \\ %1.0  
5532.232 &  0.609 &  7.696 $\pm$ 0.031 & 7.369 $\pm$ 0.028 & --                & --                \\ %0.5
5532.298 &  0.609 &  7.652 $\pm$ 0.046 & 7.343 $\pm$ 0.030 & --                & --                \\ %1.0  
5532.364 &  0.609 &  7.671 $\pm$ 0.034 & 7.446 $\pm$ 0.022 & --                & --                \\ %0.5  
5532.430 &  0.610 &  7.622 $\pm$ 0.033 & 7.263 $\pm$ 0.027 & --                & --                \\ %0.5  
5532.430 &  0.610 &  7.735 $\pm$ 0.044 & 7.384 $\pm$ 0.027 & --                & --                \\ %1.0  
5532.496 &  0.610 &  7.641 $\pm$ 0.029 & 7.368 $\pm$ 0.023 & --                & --                \\ %1.0  
5532.629 &  0.610 &  7.623 $\pm$ 0.058 & 7.385 $\pm$ 0.022 & --                & --                \\ %1.0  
5532.761 &  0.611 &  7.696 $\pm$ 0.034 & 7.360 $\pm$ 0.019 & --                & --                \\ %1.0  
5532.893 &  0.612 &  7.661 $\pm$ 0.025 & 7.384 $\pm$ 0.015 & --                & --                \\ %1.0
\hline
\multicolumn{6}{p{12.5cm}}{$^a$Pulsation phase according to ephemeris: JD$_{\rm{max}}$= 2457810.0 + 242.53 $\times$ E}\\ 
}

\end{appendix}

\end{document}